\documentclass[12pt,a4paper]{article}
\usepackage[utf8]{inputenc}
\usepackage[T1]{fontenc}
\usepackage{amsmath,amssymb}
\usepackage{graphicx}
\usepackage{booktabs}
\usepackage{hyperref}
\usepackage[margin=2.2cm]{geometry}
\usepackage{authblk}
\usepackage{cite}
\usepackage{setspace}
\title{\textbf{The Case for Space-Based Particle Colliders:\\ Orbital Infrastructure as a Path to\\Grand Unification Energy Scales}}

\author[1]{Viktor Danchev}
\affil[1]{EnduroSat}
\author[1]{Alex Dyer}
\author[1]{Sebastian Grau}
\author[1]{Guillaume Vazeille}

\date{\today}

\begin{document}
\maketitle

\begin{abstract}
The Standard Model of particle Physics has been validated to extraordinary high precision by the Large Hadron Collider (LHC).
Yet it leaves some of the most fundamental questions in Physics unresolved: the nature of dark matter, the hierarchy problem, and the unification of forces. 
Multiple next-generation terrestrial colliders have been proposed such as the Future Circular Collider (FCC) which will reach centre-of-mass energies of $\sim$100~TeV, yet the energy scales at which hints of Grand Unified Theories (GUTs) and string theory are expected to be observed ($10^{11}$--$10^{13}$~TeV) remain orders of magnitude beyond the reach of any terrestrial facility physicists even dare dream of. 
We argue that the path to these energy frontiers inevitably leads to Space. 
By examining the fundamental scaling law $E_\mathrm{cm} \propto B \times r$ for circular proton colliders, we establish that colliders of radius $10^3$--$10^5$~km are required to enter the PeV--EeV regime. 
In addition, Space-based colliders benefit from virtually free ultra-high vacuum ($<10^{10}$~particles/m$^3$ above 1000~km altitude), passive cryogenic cooling, reduction of geological and political siting constraints, and perhaps most importantly -- the substantial reduction of the thermodynamic penalty that dominates terrestrial cryogenic power budgets. 
We survey existing proposals for beyond-Earth colliders, derive order-of-magnitude requirements for an orbital collider constellation, and assess feasibility against current and near-term spacecraft capabilities in formation flying, power generation, and precision attitude control. 
We conclude that recent developments in orbital infrastructure---particularly gigawatt-scale orbital power architectures being developed for Space-based data centers---are converging with the needs of a Space-based mega collider, making serious feasibility studies warranted and promising a more certain path towards the core questions of modern Physics.
\end{abstract}

\noindent\textbf{Keywords:} particle colliders, grand unification, string theory, Space infrastructure, precision formation flying, orbital data centers

\section{The Roadblock in Fundamental Physics}

The Standard Model of particle Physics stands as one of the most precisely tested theories in the history of science. 
The discovery of the Higgs boson at the LHC in 2012 by the ATLAS and CMS collaborations~\cite{ATLAS2012,CMS2012} completed the particle content predicted by the model. 
In the decade since, the LHC---operating at a world-record centre-of-mass energy of $\sqrt{s} = 13.6$~TeV in Run~3~\cite{Evans2008,CMSRun3}---has found no statistically significant deviations from Standard Model predictions. 
Searches for supersymmetric particles, extra dimensions, and other beyond-Standard-Model (BSM) phenomena have yielded only exclusion limits~\cite{LHCExtraDim}.

Yet the Standard Model is obviously incomplete. 
It does not incorporate gravity, does not explain dark matter or dark energy (which together constitute $\sim$95\% of the universe's energy content), offers no mechanism for the matter--antimatter asymmetry, leaves neutrino masses unexplained, and provides no fundamental rationale for its own structure~\cite{PDG_GUT}. 
The \emph{hierarchy problem}---why the Higgs mass is $10^{16}$ times smaller than the Planck scale---remains perhaps the deepest puzzle~\cite{Strassler_hierarchy}.
Beyond this -- the very way in which the Standard Model is constructed remains at its core empirical recipe rather than profound mathematical derivation.

Grand Unified Theories (GUTs) propose that the strong, weak, and electromagnetic forces unify into a single gauge interaction at an energy scale $M_\mathrm{GUT} \sim 10^{14}$--$10^{16}$~GeV~\cite{PDG_GUT,GUT_Frontiers}. 
In supersymmetric extensions of the Standard Model, the three gauge coupling constants converge with remarkable precision at $M_\mathrm{GUT} \approx 2 \times 10^{16}$~GeV~\cite{Langacker2012}. 
String theory, the leading candidate for a quantum theory of gravity, predicts new Physics at scales that depend on the compactification geometry: from the Planck scale ($\sim 10^{19}$~GeV) in traditional scenarios~\cite{StringTheoryWiki}, to an intermediate scale ($\sim 10^{11}$~GeV) in Large Volume Scenario compactifications~\cite{LVS_Conlon,Burgess1999}, and potentially as low as the TeV scale in models with large extra dimensions~\cite{ADD1998,RandallSundrum1999}.

The strongest indirect constraints on the GUT scale come from proton decay searches. 
Super-Kamiokande has established lower limits on the proton partial lifetime of $\tau_p / \mathrm{Br}(p \to e^+ \pi^0) > 2.4 \times 10^{34}$~years~\cite{SuperK_proton}, ruling out minimal SU(5) and constraining SUSY GUT models. 
Next-generation experiments---Hyper-Kamiokande, DUNE, and JUNO---will improve sensitivity by an order of magnitude~\cite{HyperK_DUNE}. 
However, proton decay is an indirect probe: it constrains the GUT scale but cannot map the spectrum of new particles or determine the gauge group structure. 
Only a collider operating at or near the unification scale could provide direct experimental access to the Physics of grand unification.

The current state of Physics is a theoretically rich landscape of models with no experimental guidance as to which (if any) is correct due to the enormous gap in scales.
Worse yet -- most of these theories are so broad, that they can be effectively tuned to match any observation at the scales explored today, or the scales realistically considered in any near-term terrestrial collider. 
As Pauli would put it "They are not even wrong".


\section{From TeV to GUT Scales}

\subsection{The fundamental scaling law}

For an ultrarelativistic proton ($E \gg m_p c^2$) in a circular synchrotron, the beam energy is determined by the magnetic rigidity relation~\cite{Wiedemann2015,Holzer2013}:
\begin{equation}
E_\mathrm{beam} \approx 0.3 \, f_\mathrm{dip} \, B \, r,
\label{eq:scaling}
\end{equation}
where $E_\mathrm{beam}$ is in GeV, $B$ is the dipole magnetic field in T, $r$ is the ring radius in meters, and $f_\mathrm{dip}$ is the dipole filling factor (the fraction of the circumference occupied by bending magnets), which is typically in the range 0.6--0.8 for most colliders. 
For a head-on proton--proton collider, $E_\mathrm{cm} = 2 E_\mathrm{beam}$.

This relation has been validated across decades of collider construction. 
The LHC ($r \cong 4.25$~km, $B = 8.33$~T, $f_\mathrm{dip} \approx 0.66$) achieves $E_\mathrm{beam} \approx 7$~TeV, yielding $E_\mathrm{cm} = 14$~TeV by design~\cite{Evans2008,LHCDesignReport}. 
The proposed FCC-hh ($r \cong 14.48$~km, $B = 16$~T) targets $E_\mathrm{cm} \approx 100$~TeV~\cite{FCC_Benedikt,FCC_Feasibility}. 
The cancelled Superconducting Super Collider ($r \cong 13.85$~km, $B = 6.6$~T) was designed for $E_\mathrm{cm} = 40$~TeV~\cite{SSC}.

\subsection{Energy targets and required collider radius}

Using Eq.~\ref{eq:scaling}, we estimate the radius required to reach three target energy scales under three magnetic field assumptions (Table~\ref{tab:scaling}).

\begin{table}[h]
\centering
\caption{Required collider radius for target centre-of-mass energies under five magnetic field scenarios. $f_\mathrm{dip} = 0.7$ assumed throughout.}
\label{tab:scaling}
\begin{tabular}{@{}lcccc@{}}
\toprule
Target $E_\mathrm{cm}$ & $B = 2$~T & $B = 8$~T & $B = 16$~T & $B = 20$~T \\
 & (passive MgB$_2$) & (LHC-class) & (FCC-class) & (next-gen HTS) \\
\midrule
1~PeV ($10^6$~GeV) & 1\,190~km & 298~km & 149~km & 120~km \\
100~PeV ($10^8$~GeV) & 119\,048~km & 29\,762~km & 14\,881~km & 11\,905~km \\
$10^{14}$~GeV (non-SUSY GUT) & 795~AU & 199~AU & 100~AU & 80~AU \\
\midrule
$10^{16}$~GeV (SUSY GUT) & 79\,547~AU & 19\,920~AU & 9\,960~AU & 7\,955~AU \\
\bottomrule
\end{tabular}
\end{table}

It is clear that the GUT scale ($10^{16}$~GeV) requires colliders of interstellar dimensions ($\sim 10^{3}$--$10^{5}$~AU), consistent with earlier estimates~\cite{Siegel_BigThink}. 
However, the PeV--EeV range ($10^6$--$10^8$~GeV) requires radii of $10^2$--$10^5$~km---scales achievable with infrastructure in Earth orbit.

\begin{figure}[ht]
\centering
\includegraphics[width=\textwidth]{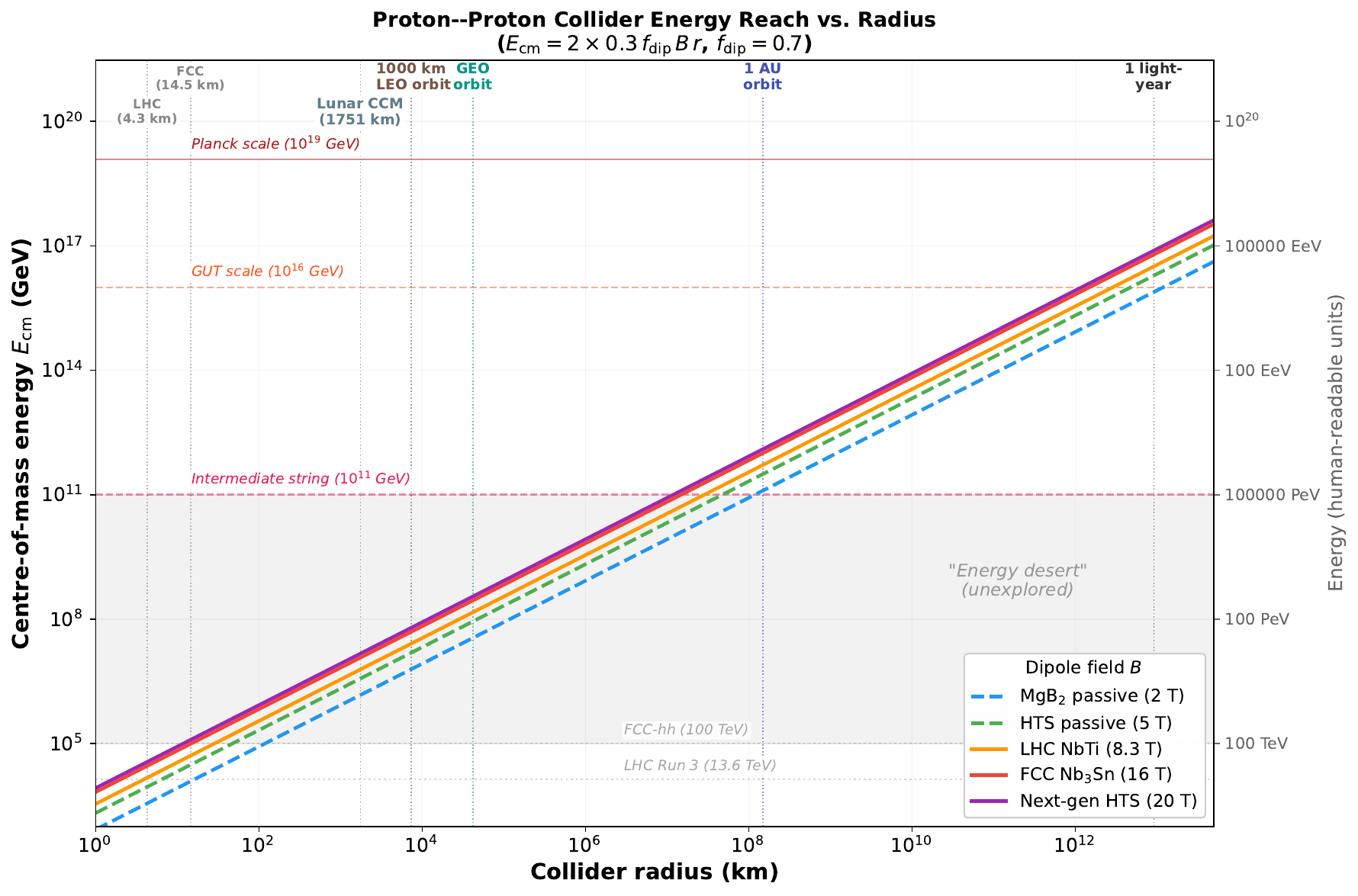}
\caption{Centre-of-mass energy reach of a proton--proton collider as a function of its radius, for dipole filling factor $f_\mathrm{dip} = 0.7$ and five magnetic field strengths spanning passively-cooled MgB$_2$ (2~T) to next-generation HTS (20~T). Vertical dashed lines indicate the radii corresponding to existing/proposed colliders and key orbital scales. Horizontal lines indicate the energy scales of the LHC, FCC-hh, intermediate string theory, GUT unification, and the Planck scale. The shaded region marks the ``energy desert'' that no current or proposed terrestrial collider can explore.}
\label{fig:ecm_scaling}
\end{figure}

The relationship between collider radius and achievable centre-of-mass energy is shown in Fig.~\ref{fig:ecm_scaling} for several magnetic field strengths, with key orbital scales and Physics energy thresholds indicated.

Based on the existing theoretical framework, there are three scenarios which must be considered:

\textbf{Optimistic:} In large extra dimension models~\cite{ADD1998,RandallSundrum1999}, the string scale could be as low as a few TeV, potentially accessible to the FCC. 
LHC searches have already excluded Kaluza--Klein gravitons below $\sim$4--5~TeV~\cite{LHCExtraDim}, and universal extra dimension compactification scales below $\sim$1.4~TeV~\cite{MUED_LHC}. 
If extra dimensions exist at the multi-TeV scale, the HL-LHC or FCC could find evidence.

\textbf{Realistic:} The Large Volume Scenario of string compactification places the string scale at $M_s \sim 10^{11}$~GeV~\cite{LVS_Conlon,Burgess1999,King2001}, motivating orbital colliders or even larger scale colliders radius of $\sim$0.1-1~AU in cislunar or inner Solar System scales.
One must not forget that the ``energy desert'' between the electroweak scale ($\sim$100~GeV) and the GUT scale ($\sim 10^{16}$~GeV) is an assumption, not an observation. 
New Physics could appear at any intermediate scale, but the higher the scale we can probe - the more likely we are to find it.

\textbf{Pessimistic:} If the traditional string/Planck scenario holds ($M_s \sim 10^{19}$~GeV), direct collider access requires radii of $\sim$100~light-years~\cite{Siegel_BigThink,Beacham2022}---beyond foreseeable technology even in the context of this work.

\section{Existing Proposals and the Technical Case for Space}

\subsection{The collider roadmap}

The near-term roadmap for hadron colliders is well-defined. 
The LHC will operate through the HL-LHC programme until $\sim$2041~\cite{Evans2008}. 
The FCC feasibility study, completed in 2025 by over 1500 physicists and engineers, found no technical showstoppers for a collider with radius of 14.5~km operating at 100~TeV, at an estimated cost of CHF~15.3 billion for the first (lepton) stage~\cite{FCC_Feasibility,FCC_SciAm}. 
China's CEPC/SppC programme targets similar parameters~\cite{CEPC}. 
The Snowmass'21 Implementation Task Force has evaluated all major proposals~\cite{Snowmass_ITF}.

Beyond these, three concepts push towards larger scales:

\begin{enumerate}
\item \textbf{Collider in the Sea (CitS):} McIntyre et al.\ propose a 320~km radius ring at 100~m depth in the Gulf of Mexico, reaching $E_\mathrm{cm} \sim 500$~TeV using 3.2~T dipoles~\cite{CitS}.
\item \textbf{Circular Collider on the Moon (CCM):} Beacham and Zimmermann sketch a hadron collider around a lunar great circle ($r \approx 1\,737$~km), reaching $E_\mathrm{cm} \sim 14$~PeV with 20~T dipoles~\cite{Beacham2022}.
\item \textbf{Solar-system scale concepts:} Fermi's ``Fermitron'' (equatorial Earth ring) and Siegel's analysis of GUT-scale requirements~\cite{Siegel_BigThink} frame the ultimate ambition.
\end{enumerate}

Notably, no published feasibility study addresses seriously an \emph{orbital} collider which can be achieved through a formation-flying satellite constellation forming a synchrotron ring in Earth (or Sun) orbit.
Such concepts have remained constrained to the realm of Sci-Fi, but we argue that with today's technology, they finally merit serious scientific and engineering consideration.

\subsection{Technical advantages of a Space-based collider}

While Space is much more difficult to operate in, orbital colliders offers several order-of-magnitude advantages over terrestrial and lunar alternatives which merit the consideration:

\textbf{Free ultra-high vacuum:} 
The LHC beam vacuum specification requires a hydrogen-equivalent molecular density below $\sim 10^{15}$~molecules/m$^3$ in the arcs and $\sim 10^{13}$~molecules/m$^3$ near interaction points, corresponding to pressures of $10^{-7}$--$10^{-9}$~Pa~\cite{LHC_vacuum_Grobner,LHC_vacuum_Bregliozzi}. 
This is achieved through 54~km of ultra-high vacuum piping, 250\,000 welded joints, 18\,000 vacuum seals, and extensive cryopumping~\cite{CERN_vacuum}. 
In contrast, the natural particle density above 1000~km altitude in Earth orbit is $\sim 10^{10}$--$10^{11}$~molecules/m$^3$, and above 10\,000~km altitude (MEO) it falls to $\sim 10^{7}$--$10^{8}$~molecules/m$^3$~\cite{NRLMSISE,USStdAtm1976}. 
At geostationary orbit ($\sim$36\,000~km altitude), densities of $\sim 10^{5}$~molecules/m$^3$ are 7--8 orders of magnitude better than the LHC beam pipe---requiring zero additional infrastructure.

A direct density comparison cannot be made, however, due to the differing gas composition. 
The LHC specification is driven by nuclear beam-gas scattering, which limits beam lifetime~\cite{LHC_vacuum_Grobner}. 
The beam lifetime from this process is $\tau \approx 1/(n\,\sigma_\mathrm{nuc}\,c)$, where $n$ is the residual gas number density, $\sigma_\mathrm{nuc}$ is the nuclear interaction cross-section, and $c$ is the speed of light. 
In the LHC, the residual gas is predominantly H$_2$, with $\sigma_\mathrm{nuc}(p\text{--H}_2) \approx 50$~mb. 
At 1000~km altitude, the residual atmosphere is dominated by atomic Helium (A$=$4 worst case), for which $\sigma_\mathrm{nuc}(p\text{--He}) \approx A^{2/3} \times \sigma_{pp} \approx 126$~mb~\cite{PDG_cross}. 
Despite of this $\sim$2.5$\times$ larger cross-section per atom, the density of $n \sim 10^{10}$~m$^{-3}$ at 1000~km yields a beam-gas lifetime of:
\begin{equation}
\tau_{1000\,\mathrm{km}} \approx \frac{1}{10^{10} \times 1.26\times 10^{-29} \times 3\times 10^{8}} \approx 2.5 \times 10^{10}~\mathrm{s} \approx 850~\mathrm{years}.
\end{equation}
In GEO orbital height, where the residual gas is predominantly atomic hydrogen at $n \sim 10^{5}$~m$^{-3}$, the lifetime exceeds $10^{8}$~years. 
Both values significantly exceed LHC's 100-hour design lifetime, confirming that the natural vacuum of Space is more than adequate even when species-dependent cross-sections are properly accounted for. 
We note that the charged particle environment (Van Allen belt protons, solar wind) requires separate analysis, as beam--plasma interactions differ qualitatively from beam--gas scattering; however, the number densities of trapped belt protons ($\sim 10^{7}$--$10^{8}$~m$^{-3}$) are still far below the LHC gas density threshold.

\textbf{Synchrotron radiation disposal:} In the LHC, synchrotron radiation ($\sim$3.6~kW per beam at 7~TeV~\cite{Evans_SLAC}) is intercepted by a beam screen at 5--20~K inside the 1.9~K cold bore. 
Removing this heat at 1.9~K costs $\sim$700~W of electrical power per watt of heat, due to the Carnot penalty~\cite{LHCDesignReport}. 
This cryogenic overhead dominates LHC's $\sim$120~MW power budget. 
For the FCC-hh, synchrotron radiation rises to $\sim$5~MW total~\cite{Shiltsev_ModFut}. 
In Space, synchrotron photons can simply escape into the vacuum, as noted for the lunar CCM concept where ``open plane dipole magnets would allow the photons to escape into the Moon sky''~\cite{Beacham2022}. 
This eliminates the single largest power cost of a terrestrial superconducting collider.

However, at PeV beam energies the total synchrotron radiation power becomes significant and must be quantitatively assessed. 
The energy loss per revolution for an ultrarelativistic proton ($\beta \approx 1$) of energy $E = \gamma m_p c^2$ in dipole magnets with bending radius $\rho = r\,f_{\mathrm{dip}}$ is (Wiedemann~\cite{Wiedemann2015}, Ch.~24, Eq.~24.11):
\begin{equation}
U_0 = \frac{q^2 \gamma^4}{3\varepsilon_0 \,\rho},
\label{eq:U0}
\end{equation}
where $q$ is the proton charge and $\varepsilon_0$ the permittivity of free Space. 
For the LHC at 7~TeV ($\gamma = 7471$, $\rho = 2800$~m), Eq.~\ref{eq:U0} gives $U_0 \approx 6.7$~keV per turn, consistent with the design value~\cite{Evans_SLAC}. 
The total synchrotron radiation power for two beams, each carrying current $I_\mathrm{beam}$, is:
\begin{equation}
P_\mathrm{SR} = 2\,\frac{U_0 \, I_\mathrm{beam}}{e}.
\label{eq:Psr}
\end{equation}
The critical photon energy, which characterises the spectral hardness, is
\begin{equation}
\varepsilon_c = \tfrac{3}{2}\hbar c\,\gamma^3/\rho.
\label{eq:spH}
\end{equation}
Table~\ref{tab:synrad} evaluates Eqs.~\ref{eq:U0}--\ref{eq:spH} for representative configurations considering a LEO and a GEO collider ring used a their maximum beam energy (FP = Full Power) and at a fraction of it (LP = Low Power).

\begin{table}[h]
\centering
\caption{Synchrotron radiation parameters for representative collider configurations. LHC and FCC-hh use their actual design parameters; Orbital collider entries assume $f_\mathrm{dip} = 0.7$, $B = 2$~T, $I_\mathrm{beam} = 0.58$~A.}
\label{tab:synrad}
\begin{tabular}{@{}lccccccc@{}}
\toprule
Configuration & $r$ & $\rho$ & $B$ & $E_\mathrm{cm}$ & $U_0$ & $\varepsilon_c$ & $P_\mathrm{SR}$ \\
\midrule
LHC & 4.25~km & 2\,804~m & 8.3~T & 14~TeV & 6.7~keV & 44~eV & 7.8~kW \\
FCC-hh & 14.5~km & 10\,417~m & 16~T & 100~TeV & 4.7~MeV & 4.3~keV & 4.7~MW \\
LEO~ring (LP) & 7\,378~km & 5\,164\,600~m & 2~T & 500~TeV & 5.9~MeV & 1.1~keV & 6.9~MW \\
LEO~ring (FP) & 7\,378~km & 5\,164\,600~m & 2~T & 6.2~PeV & 140~GeV & 2.1~MeV & 162~GW \\
GEO~ring (LP) & 42\,146~km & $2.95\times 10^7$~m & 2~T & 1~PeV & 16.6~MeV & 1.5~keV & 19~MW \\
GEO~ring (FP) & 42\,146~km & $2.95\times 10^7$~m & 2~T & 35.4~PeV & 26~PeV & 67~MeV & 30~PW \\
\bottomrule
\end{tabular}
\end{table}

Table~\ref{tab:synrad} reveals a critical design insight: at a given beam energy, synchrotron radiation power scales as $P_\mathrm{SR} \propto B \times E_\mathrm{cm}^3$ (since $P_\mathrm{SR} \propto E^4/\rho$ and $\rho = E/(0.3\,B)$). 
Lower magnetic fields therefore produce \emph{less} synchrotron radiation at the same energy, at the cost of a proportionally larger ring. 
The LEO ring configuration ($r = 7378$~km corresponding to 1000 km altitude orbit) reaches $E_\mathrm{cm} = 500$~TeV---5$\times$ the FCC-hh---with a total synchrotron radiation power comparable to the FCC-hh's load and well within the power envelope of gigawatt-scale orbital architectures discussed in Section~4.4. 

Crucially, although the synchrotron radiation escapes freely in an open-geometry magnet design, it is emitted tangentially to the beam orbit in a narrow cone of opening angle $\sim 1/\gamma$ and will illuminate neighbouring satellite modules at grazing incidence. 
At the critical photon energies of $\cong1$~keV shown in Table~\ref{tab:synrad}, this can pose radiation damage risk to electronics and structural components.
This is a meaningful reason to increase the spacing between satellites, sacrificing the maximum energy available to the accelerator linearly in favour of reducing direct additional beam damage.
For the low power example in the LEO ring configuration, $\gamma \cong 266841$, so the emission will be with an angle of the cone $\cong 0.77$ arcsec.
This value implies less than approximately 1.7 million satellites, which would likely be around the practical limit in all cases.
A detailed radiation environment analysis, accounting for the angular distribution of synchrotron photons and the geometry of the satellite constellation, is required and is identified as a step for future study.
Nevertheless, given the general radiation hardening of spacecraft components or at least radiation tolerance in NewSpace systems, the radiation contribution outside of the main beam is not expected to influence the constellation or other satellites significantly.

\textbf{Passive cryogenic cooling:} The JWST demonstrates passive cooling to $\sim$40~K at the Sun--Earth L2 point using a sunshield~\cite{JWST}. 
High-temperature superconducting (HTS) magnets operating at 20--40~K could potentially be cooled passively or with greatly reduced cryogenic systems in the radiative environment of Space, where the cosmic microwave background provides a 2.7~K thermal sink.
Low Earth Orbit (LEO) is not the same as a Lagrange point in terms of the passive cooling benefits, but certain orbits such as Dusk-Dawn Sun-Synchronous orbits (SSO) can provide nearly all of the benefits such as constant Sun/Earth angle and constant Deep Space sink visibility, albeit with much larger IR flux coming from the Earth.  

\textbf{No geological or political siting constraints:} The FCC requires a 91~km long tunnel through the geology of the Geneva basin, facing opposition from landowners and environmental groups~\cite{FCC_SciAm}. 
The CCM requires lunar tunnel boring or surface construction~\cite{Beacham2022}.
An orbital collider avoids all terrestrial siting issues entirely.
While the LEO environment is already quite full, orbits beyond 1000 km and especially beyond GEO are rather unpopulated and will benefit both from the larger radius, better vacuum, absence of van Allen belts particles, etc.

\subsection{Power scaling analysis}

During LHC operation, the CERN site draws $\sim$200~MW peak power~\cite{CERN_power,IEEE_LHC_power}. 
Out of this, the LHC machine consumes $\sim$120~MW, dominated by cryogenics ($\sim$28--35~MW), cooling and ventilation ($\sim$25--30~MW), and magnet power converters ($\sim$15--20~MW)~\cite{CERN_elec_2019}. 
The RF system delivering power to the beam draws only $\sim$1--2~MW, and the actual beam power (replenishing synchrotron radiation) is only at $\sim$7~kW~\cite{Evans_SLAC,CERN_RF}. 
The ratio of consumed electrical power to beam power is on the order of $\sim$17\,000, reflecting the thermodynamic cost of maintaining a cryogenic, vacuum-enclosed, underground facility.

For a Space-based collider, the dominant power scaling laws are:
\begin{itemize}
\item Synchrotron radiation: $P_\mathrm{SR} \propto E^4 / r \propto B^4 r^3$ (at constant $B$)~\cite{Wiedemann2015}
\item Static cryogenic load: $\propto r$ (proportional to magnet count and thus circumference/radius)
\item RF beam power: $\propto P_\mathrm{SR}$ (must replenish radiation losses)
\item Vacuum: zero in Space (vs.\ $\propto r$ terrestrially)
\end{itemize}

The elimination of the cryogenic Carnot penalty and the needed vacuum systems could reduce the overhead factor by one to two orders of magnitude, making very large rings dramatically more power-efficient per unit of Physics to reach compared to terrestrial equivalents.

\subsection{The low-field trade-off: passive cooling vs.\ satellite count}

A Space-based collider opens an option unavailable to terrestrial machines: the use of low-field superconducting magnets that can be cooled mostly or even entirely by passive radiation to the cold of Space, thus significantly reducing or even eliminating active cryogenic systems altogether. 
Materials such as MgB$_2$ achieve superconductivity at $\sim$39~K and can sustain fields of 2--4~T~\cite{MgB2}, while REBCO (rare-earth barium copper oxide) tapes operate at 40--77~K with fields of 5--10~T~\cite{REBCO_magnets}. 
Both temperature ranges are accessible through passive radiative cooling with sunshields, as demonstrated by JWST's 40~K operating temperature at L2~\cite{JWST}.
While JWST benefits from a location at the L1 point, any orbital ring which maintains local time of the ascending node close to 6:00 and is far enough from the surface of the Earth to avoid its radiative heating, can achieve similar cooling capabilities.

The trade-off is governed by two competing scaling relations. For a fixed target energy, Eq.~\ref{eq:scaling} gives $r \propto 1/B$: halving the magnetic field doubles the required radius of the collider constellation ring.
Since the number of satellite modules scales as $N_\mathrm{sat} = r / L_\mathrm{cell}$ (where $L_\mathrm{cell}$ is the FODO cell length, set by beam optics), we have:
\begin{equation}
N_\mathrm{sat} \propto \frac{E_\mathrm{cm}}{B \cdot L_\mathrm{cell}}
\label{eq:nsat}
\end{equation}
At fixed $E_\mathrm{cm}$ and $L_\mathrm{cell}$, the satellite count scales as $N_\mathrm{sat} \propto 1/B$. 
A collider using 2~T passively-cooled MgB$_2$ magnets requires 10$\times$ more satellites than one using 20~T actively-cooled HTS magnets for the same energy reach.

However, the cost per satellite in the constellation is not constant with respect to $B$. 
A passively-cooled 2~T magnet module requires no cryoplant, no helium supply chain, no active thermal control beyond a sunshield, and can use commercially mature superconducting wire. 
The mass, complexity, and power consumption per satellite unit will be substantially lower. 
If we define $\mathcal{C}(B)$ as the cost per satellite module which is some non-linear function dependent on the magnetic field which a given FODO cell satellite can produce, the total system cost scales as:
\begin{equation}
\mathcal{C}_\mathrm{total} \propto \frac{E_\mathrm{cm}}{B \cdot L_\mathrm{cell}} \times \mathcal{C}(B)
\end{equation}
There exists some optimum field strength $B^*$ that minimises $\mathcal{C}_\mathrm{total}$. 
If the per-unit satellite cost scales faster than linearly with $B$ (due to the exponentially increasing difficulty of high-field magnet construction and the need for active cryogenics above $\sim$10~T), then $B^*$ would likely be within the lower fields -- likely within the range 2--5~T assuming that passive cooling is possible. 
This is the very different from the terrestrial situation, where the fixed cost of civil engineering (tunnelling) makes maximising $B$ the dominant strategy. 
The detailed determination of $B^*$ for a Space-based architecture, accounting for launch costs, module mass, and orbital deployment logistics, is an important subject for further research based on a preliminary FODO satellite design concept.

\section{Feasibility and Spacecraft Requirements}

\subsection{Constellation architecture}

The LHC beam optics are based on the FODO lattice: each element consisting of a unit cell with length $\sim$107~m containing six dipole magnets and two quadrupole magnets~\cite{LHCDesignReport,Holzer2013,Giovannozzi2022}. 
The 1232 main dipoles and $\sim$474 quadrupoles form a discrete polygon, not a continuous arc~\cite{Evans2008}. 
An orbital collider would replicate this architecture with each FODO cell implemented as a satellite module or small cluster.

For a ring of circumference $C = 47\,000$~km (roughly 1000 km LEO altitude scale) with optimised 1~km distance between the cells, one would require:
\begin{itemize}
\item $\sim$47\,000 FODO cells, each of which would be a satellite
\item $\sim$470\,000--560\,000 dipole magnets (10--12 per satellite cell)
\item $\sim$94\,000 quadrupole magnets
\end{itemize}

Assuming the lower power regime of $E_\mathrm{cm} \approx 1.0$~PeV at $B = 2$~T and $f_\mathrm{dip} = 0.7$, this would imply individual satellite cell size of ~700 m.
There is another trade-off here: between the size of the individual satellite FODO cell and the number of satellites.
Methods such as in-orbit assembly can be considered to assemble longer FODO cells by combining separate magnet sections and forming larger cells.
Alternatively, smaller individual satellites can be used either reducing the $f_\mathrm{dip}$ value (and respectively the maximum reachable energy scale), or simply increasing the number of satellites.
Similarly to the magnetic field trade-off - there is some optimal number of satellites which can be found based on trade-offs between complexity for in-orbit assembly of a smaller number of satellites and the simplicity of launching more satellites at a lower unit cost.
Since $f_\mathrm{dip}$ enters Eq.~\ref{eq:U0} linearly through $\rho$, reducing the size of a cell in the above example to 100 meters for example would imply total power increase of 7 times relative to Table~\ref{tab:synrad} and 7 times higher mean energy for the escaping radiation.

An individual FODO cell satellite would be a long tube with all the required dipole and quadrupole magnetic optics around a hollow structure and attached solar panels and radiators.
Given that the largest rocket fairings have a diameter of ~10 m and height of ~15-20 meters, it is safe to assume that each FODO cell will be assembled of several individual satellites or that the individual satellite will deploy large booms between the magnet optics, reconfiguring itself in orbit to achieve the desired geometry from a much more compact initial state.
This sets practical FODO cell size on the order of 30-200 meters, which implies the required number of satellites to be from a few hundred thousands to approximately 1 million depending on the final design.
While large, this satellite count is within an order of magnitude of planned mega-constellations.
SpaceX Starlink targets $\sim$42\,000 satellites with a significant portion already in Space.
More recently, SpaceX \href{https://fccprod.servicenowservices.com/icfs?id=ibfs_application_summary&number=SAT-LOA-20260108-00016}{filed for up to 1 million satellites} for orbital data centers in Space, where the modular, mass-manufactured architecture within a single orbital plane is conceptually the same.
We will explore additional synergies between the Space data centers application and technologies required for an orbital collider, notably large deployable structures.

\subsection{Pointing and formation flying}

The beam aperture in the LHC is $\sim$30~mm radius. 
For satellites spaced $\sim$1~km apart, maintaining the beam within this aperture requires relative attitude control in the direction transverse to velocity to $\sim\pm$6~arcsec and relative formation on the centimeter scale. 
This tolerance applies to quasi-static alignment.
While beam optics are sensitive to both static misalignment and dynamic jitter, the relevant timescales differ. 
Static alignment must hold over the orbital perturbation timescale (hours to days), while dynamic jitter must remain small compared to the beam size over the betatron oscillation period of the beam. 
For a LEO-scale ring with revolution frequency $f_\mathrm{rev} \approx c/C \approx 6.5$~Hz and betatron tune $Q \approx 100$, the betatron frequency is $\sim$650~Hz, but the magnet alignment does not have to be tracking this frequency -- the beam's own transverse oscillations are handled by the FODO optics. 
What matters is that the magnetic centre of each element remains within the beam aperture on timescales set by orbital perturbations, which evolve over minutes to hours.

The most relevant flight demonstration is ESA's PROBA-3, launched in December 2024, which achieved autonomous formation flying between two spacecraft at 150~m separation with millimetre-level longitudinal accuracy and sub-millimetre lateral accuracy, maintained for several hours without ground control~\cite{PROBA3_ESA,PROBA3_wiki}. 
This was demonstrated at apogee altitudes exceeding 50\,000~km, in a gravitational environment directly comparable to MEO/GEO orbits. 
The PROBA-3 lateral accuracy of $<$1~mm at 150~m separation corresponds to $\sim$1.4~arcseconds, substantially better than the $\sim$6~arcsecond tolerance derived above. 
Scaling from 150~m to 1~km baselines (which would scale linearly in this regime, assuming corresponding sensor sensitivity scaling), the same angular precision would yield $\sim$7~mm positioning accuracy -- well within the required $\pm$30~mm beam aperture.

Additional relevant demonstrations include: the PRISMA mission, which achieved cm-level autonomous relative navigation in LEO~\cite{PRISMA_DAmico}; and the GRACE Follow-On laser ranging interferometer, which provides nanometre-level inter-satellite ranging over $\sim$220~km~\cite{GRACE_FO}.

The principal perturbation forces driving differential displacement between adjacent FODO satellites at upper LEO and GEO are solar radiation pressure (SRP) and lunisolar gravitational gradients.
For equivalent FODO cell satellites pointing in the same direction (for example 6:00 LTAN ring), the area-to-mass ratio difference is expected to be negligible.
For area-to-mass ratio delta on the order of $\Delta(A/m) \sim 10^{-3}$~m$^2$/kg (considering surface variations, minor pointing misalignments, etc.), the differential SRP acceleration is $\sim 5 \times 10^{-10}$~m/s$^2$. 
Over a 1000~s correction cycle, this produces a differential drift of $\sim$0.25~mm---well within tolerance and correctable by mN-class electric propulsion systems on the market.
Gravitational gradient forces at 1000 km LEO orbit for a 1~km baseline are on the order of $\Delta a \sim 2GM_\oplus \Delta r / r^3 \approx 10^{-9}$~m/s$^2$, producing comparable drifts. 
Of course, the various Earth non-spherical harmonic geopotential terms are signiificantly stronger at 1000 km, but all of these gravitationally-arising perturbations are deterministic and easily predictable, enabling feedforward compensation.
We conclude that the formation-keeping requirements, while demanding, are within the demonstrated state of the art for precision formation flying, and do not require extrapolation beyond current capabilities.

It is worth to mention that the satellites quoted above are flagship missions with significant budget, while the scale of this concept would require one to consider NewSpace mentality and much more affordable technology.
Nevertheless, recent trends in NewSpace capabilities provide significant reason for optimism that such performance can be achieved at scale.
Appropriate fiducials and multi DoF mN class thrusters available on each FODO cell, as well as inter-satellite link for real-time tasking across the constellation would enable a feasible fully autonomous feedforward control.

\subsection{Luminosity considerations}

A collider is scientifically useful only if it delivers adequate luminosity to observe the processes of interest. The peak luminosity of a circular hadron collider is~\cite{Shiltsev_ModFut}:
\begin{equation}
\mathcal{L} = \frac{N_b^2\, n_b\, f_\mathrm{rev}\, \gamma}{4\pi\,\varepsilon_n\,\beta^*} \, F
\label{eq:lumi}
\end{equation}
where $N_b$ is the number of particles per bunch, $n_b$ the number of bunches, $f_\mathrm{rev}$ the revolution frequency, $\gamma$ the Lorentz factor, $\varepsilon_n$ the normalised transverse emittance, $\beta^*$ the beta function at the interaction point (IP), and $F \leq 1$ is a geometric reduction factor from the crossing angle.

For a LEO-scale ring ($C \approx 47\,000$~km), $f_\mathrm{rev} = c/C \approx 6.5$~Hz, compared to the LHC's $f_\mathrm{rev} = 11\,245$~Hz. 
This factor-of-$\sim$1\,730 reduction in revolution frequency (directly dictated by the circumference increase) is the principal luminosity challenge.
With LHC-nominal bunch parameters ($N_b = 1.15 \times 10^{11}$, $n_b = 2808$, $\varepsilon_n = 3.75~\mu$m, $\beta^* = 0.55$~m), the na\"ive luminosity estimate is $\mathcal{L} \sim 6.5 \times 10^{32}$~cm$^{-2}$s$^{-1}$ -- almost two orders of magnitude lower than the LHC design value.

However, several factors compensate this deficit partially. 
First, the Lorentz factor $\gamma$ enters linearly in Eq.~\ref{eq:lumi} and is $\sim$140$\times$ larger at 1~PeV than at 14~TeV, providing a direct luminosity gain. 
Furthermore, a very large ring can accommodate far more bunches: $n_b \propto C$, so the number of bunches scales up by $\sim$1\,730$\times$ relative to the LHC, exactly cancelling the $f_\mathrm{rev}$ reduction. 
In this limit, the luminosity per IP is determined by the total beam current and the IP optics, not the ring size:
\begin{equation}
\mathcal{L} \approx \frac{I_\mathrm{beam}^2\, \gamma}{4\pi\, e^2\, n_b\, f_\mathrm{rev}\, \varepsilon_n\, \beta^*}\, F \propto \frac{I_\mathrm{beam}^2 \, \gamma}{\varepsilon_n \, \beta^*}
\end{equation}
for fixed total beam current $I_\mathrm{beam} = N_b\, n_b\, e\, f_\mathrm{rev}$. 
Finally, synchrotron radiation damping times scale as $\tau_\mathrm{damp} \propto R^2/E^3$ at constant $B$ (where $E \propto R$), this gives $\tau_\mathrm{damp} \propto 1/R$. 
At the LHC, $\tau_\mathrm{damp} \approx 13$~hours, which is comparable to the Physics fill duration, so radiation damping has negligible impact on the beam emittance during operation. 
For a PeV-scale Space collider with much larger $R$ and $E$, the damping time shrinks to only an hour or even minutes, meaning the beam emittance is actively driven toward its equilibrium value during a fill. 
Since luminosity scales as $1/\varepsilon_n$ (Eq.~\ref{eq:lumi}), this natural emittance reduction provides a direct luminosity benefit unavailable at LHC energies.

Nevertheless, luminosity remains a significant design challenge and must be a central focus of any detailed feasibility study. 
Key open questions include: the achievable $\beta^*$ with a distributed satellite-based final-focus system, the beam--beam tune shift limit at PeV energies; the number and design of interaction points and detector stations and the expected event rates for BSM processes at PeV parton centre-of-mass energies, where partonic cross-sections for heavy new-particle production may be enhanced relative to lower energies. 
We note that the integrated luminosity requirement depends strongly on the signal cross-sections, which are unknown \emph{a priori} in the energy desert. 
An operational lifetime of several decades may be required to compensate lower instantaneous luminosity.
Given the design lifetime of typical satellite busses, this may require constellation replenishment or even lead to "hot" swaps -- additional FODO cell satellites available to rapidly reposition in case of any one-off cell replacement towards the end of its operational lifetime.
While this sounds much more demanding than terrestrial collider maintenance - similar considerations are once again parallel to the availability and service continuation for Space data centers.

\subsection{Power generation: the orbital data centre parallel}

The Space Sector has undergone a major transformation in the past 2 decades with companies like SpaceX leading the way in a transformation from traditional players towards a NewSpace economy.
Perhaps the most transformative recent development relevant for the feasibility of a Space-based collider is the emergence of gigawatt-scale orbital power architectures, driven by the commercial data centre industry. 
Starcloud (formerly Lumen Orbit) has proposed a 5~GW orbital data centre with solar arrays spanning 4~km $\times$ 4~km~\cite{Starcloud_WP,Starcloud_NVIDIA}. 
Google's Project Suncatcher envisions solar-powered satellite constellations for AI compute, with a demonstration mission planned for 2027~\cite{Google_Suncatcher}. 
The European ASCEND study projects a pathway from 50~kW demonstrators to 1~GW orbital facilities by mid-century~\cite{ASCEND}. 
IEEE Spectrum analysis estimates a 1~GW orbital data centre would require $\sim$4\,300 satellites~\cite{IEEE_orbital_DC}, each of which is individually much more capable power-wise and thermally than any currently flying today.

These power scales are directly relevant. 
A 1\,000~km altitude LEO collider operating at $\sim$1~PeV with efficient Space-based cryogenics would require total power on the order of $\sim$1--10~GW, dominated by RF power to compensate synchrotron radiation ($P_\mathrm{SR} \propto B^4 R^3$) and distributed cryogenic systems. 
This is precisely the power regime that the orbital data centre industry is targeting for commercial deployment within the 2030s.

The convergence is not coincidental: both applications require gigawatt-scale Space-based solar power, high-efficiency thermal management in vacuum, modular satellite constellation architectures, and autonomous distributed operations. 
Investment in orbital data centre infrastructure directly enables the power, thermal, and constellation management technologies needed for a Space-based collider, leaving the development of the specific FODO cell apparatus capable of working in Space.

\subsection{Preliminary orbital considerations}

The discussion so far was focused on the vacuum conditions alone, but the choice of orbit must balance vacuum quality, thermal and power environment, radiation, required amount of satellites and orbital stability:

\begin{itemize}
\item \textbf{Lower LEO (200--600~km):} Insufficient vacuum due to atomic oxygen ram flux ($\sim 10^{14}$ atoms/cm$^2$/s on ram surfaces~\cite{AO_LEO}); orbital drag requires continuous reboosting;
\item \textbf{Upper LEO (1\,000--2\,000~km):} Excellent vacuum ($<10^{11}$~molecules/m$^3$) lower exposure to Van Allen belts and sufficient $J_2$ perturbation strength to maintain Sun-Synchronous Orbit almost passively, circumferences of ~47\,000-53\,000~km yield up to $E_\mathrm{cm} \sim $6-12~PeV;
\item \textbf{MEO (2\,000--20\,000~km):} Excellent vacuum ($<10^{8}$~molecules/m$^3$); Van Allen radiation belts are a concern for electronics but likely not for the beam itself; circumferences of 53\,000--165\,000~km yield up to $E_\mathrm{cm} \sim 18$~PeV.
\item \textbf{GEO and beyond ($>$35\,000~km):} Near-perfect vacuum ($<10^{6}$~molecules/m$^3$); the most stable orbital environment; circumference $\sim$260\,000~km yields $E_\mathrm{cm} \sim 36$~PeV. Gravitational perturbations from the Moon and Sun must be compensated but are predictable.
\end{itemize}

Regardless of the altitude and vacuum conditions, dusk-dawn LTAN ranges (close to 6:00) provide the best power and thermal results:
\begin{itemize}
    \item Constant or near-constant Sun-viewing for the entire constellation providing maximum power.
    \item Constant deep-Space view and Sun shading for radiators deployed transversely to the orbital plane (taking into account Earth's view factor).
    \item Constant solar radiation pressure across the constellation, minimizing differential SRP effects on the relative orbits. 
\end{itemize}

For altitudes above ~1000 km, however, maintaining a dusk-dawn altitude requires using propulsion as the $J_2$ perturbation alone is not sufficient to drive the full angular rate of the RAAN required to maintain the relative view of the Sun.
Satellites above this altitude must either carry significant additional fuel (which becomes impractical for GEO-equivalent SSO orbit) or will experience different relative illumination conditions for different parts of the year.
The latter would require either significantly larger solar panels to account for projections and/or multi-degrees of freedom solar array drive mechanisms with the optimal solution to be based on the final FODO cell power needs.
Furthermore, the number of satellites (and the total mass required to orbit) will inevitably also grow linearly with the orbital altitude or provide diminishing returns on the larger radius due to a smaller filling factor.
Finally, the actual launch cost per kilogram depends strongly on the orbital altitude (roughly linearly by virtue of the gravitational potential energy). 
Therefore, the upper LEO orbit of ~1,000~km appears to be the optimal location for a first generation of an orbital collider.
Once again, the parallel to orbital data centers is clear - a dusk-dawn orbit in LEO has been preferred for those architectures for nearly the same reasons \cite{Starcloud_WP,Starcloud_NVIDIA}.

\section{Conclusion}

The history of particle Physics is a history of scaling: from Lawrence's ~2~cm cyclotron to the 4.3~km LHC: each generation of collider opened new energy regimes and revealed (or rejected) new Physics. 
The Standard Model was built on discoveries made possible by this progression. 
Yet the next major energy frontier -- the PeV to EeV scale where GUT effects, intermediate-scale string Physics, or entirely unexpected phenomena may await, is unreachable by any terrestrial collider in the current or proposed roadmap. 
The FCC-hh, the most ambitious approved project, will reach 100~TeV: still $10^{11}$ times below the canonical GUT scale.

We believe that the path to these energy scales inevitably leads to Space, and that the technical case for a Space-based collider is real and feasible within our lifetimes. 
The natural vacuum above 1000~km altitude exceeds LHC beam-pipe quality by orders of magnitude, eliminating the most complex and expensive subsystem of a terrestrial collider.
Synchrotron radiation escapes freely into Space, removing the Carnot penalty that dominates terrestrial cryogenic power budgets. 
Passive cooling to temperatures compatible with HTS magnets is achievable with sunshield technology already demonstrated by JWST. 
The formation-flying precision required for beam optics is orders of magnitude less demanding than capabilities already demonstrated by missions such as PROBA-3.

Most significantly, the commercial Space industry is independently developing the key enabling technologies based on ever-increasing computing demands. 
Gigawatt-scale orbital solar power, modular satellite constellation architectures, and autonomous distributed operations are being driven by the multi-billion-dollar orbital data centre market -- not by particle Physics. 
A Space-based collider can ride this wave of investment rather than generating it, and lead to the next insights to push Physics to the frontier.
Ironically, the entire computing revolution, Moore's law and today's technological industry rides on the discoveries of Quantum Physics set in motion more than a century ago by the first colliders!

We do not claim that a Space-based collider is imminent or that all of the technical challenges are solved. 
Significant open questions remain: the radiation hardness of superconducting magnets in the Space environment, the logistics of deploying and maintaining millions of satellite modules with flagship mission-level of precision, the beam injection and extraction scheme, detector design for interaction points, and the overall cost.
What we do argue is that the alternative -- continuing to build incrementally larger terrestrial colliders, each taking decades and costing tens of billions of dollars, before ultimately concluding that expansion into Space is necessary -- is the less efficient path. 
The scaling laws are clear and the technical synergies with commercial Space infrastructure are compelling. 
Serious feasibility studies for orbital collider architectures should begin now, in parallel with the FCC programme, so that when the terrestrial roadmap reaches its practical limits, the Space-based alternative is ready.

The universe does not owe us the convenience of placing new Physics at the energy scale of our next affordable terrestrial collider. 
If fundamental unification occurs at $10^{11}$~GeV or above -- as the most theoretically motivated scenarios suggest -- then Space-based particle accelerators are not a speculative luxury but a scientific necessity to push us to the next technological revolution!

\end{document}